\documentclass[twocolumn,article,amsmath,amssymb]{revtex4}
\usepackage{graphicx}% Include figure files
\usepackage{graphics}
\usepackage{dcolumn}% Align table columns on decimal point
\usepackage{bm}% bold math
\usepackage{units}
\usepackage{braket}
\usepackage{soul}
\usepackage{xcolor}
\usepackage{amsmath}
\usepackage{xfrac}
\usepackage{caption}
\usepackage{subcaption}
\usepackage{epstopdf}
\usepackage{physics}
\usepackage{color}
\usepackage{hyperref}

\makeatletter
\newenvironment{figurehere}
{\def\@captype{figure}}
{}
\makeatother

\begin{document}

\def\e{\mathop{\rm \mbox{{\Large e}}}\nolimits}
\def\im{\mathop{\rm \od{\iota}}\nolimits}
\newcommand{\ts}[1]{\textstyle #1}
\newcommand{\bn}[1]{\mbox{\boldmath $#1$}}
\newcommand{\bc}{\begin{center}}
\newcommand{\ec}{\end{center}}
\newcommand{\be}{\begin{equation}}
\newcommand{\ee}{\end{equation}}
\newcommand{\bea}{\begin{eqnarray}}
\newcommand{\eea}{\end{eqnarray}}
\newcommand{\ba}{\begin{array}}
\newcommand{\ea}{\end{array}}
\newcommand{\red}[1]{\textcolor{black}{#1}}

\title{Thermodynamic criticality in seismicity: Uniqueness of surface-energy scaling in the fragment-asperity model}

\author{O. Sotolongo-Costa}%\footnote{(O. Sotolongo-Costa): osotolongo@gmail.mx}}

\affiliation{Instituto de Investigaci\'on en Ciencias B\'asicas y Aplicadas, Universidad Autónoma del Estado de Morelos, Av. Universidad 1001, CP 62209, Cuernavaca, Morelos, México}

\author{A. V. Mora-Rodr\'iguez}

\affiliation{Colegio Hellen Keller, Campus Ahuatl\'an, CP 62130, Cuernavaca, Morelos, M\'exico}

\author{M. E. Mora-Ramos\footnote{(M. E. Mora-Ramos): memora@uaem.mx}}

\affiliation{Centro de Investigación en Ciencias-IICBA, Universidad Autónoma del Estado de Morelos, Av. Universidad 1001, CP 62209, Cuernavaca, Morelos, México}

\begin{abstract} 
We propose that a linear relation between the energy of stress-bearing interactions and the surface of contact within the fragment-asperity model for earthquakes. It reveals as the only one that leads to a closed elementary form for a well-defined total entropy as a function of non-extensivity parameter, $q$. By writing the total Tsallis entropy as a function of $q$, a critical range of values is identified: $1.4\lesssim q\lesssim 1.8$. Such interval of $q$-values corresponds to the strong variation of entropy and contains the most of reported results for this parameter determined for main-shocks around the world in recent decades, indicating the role of $q$ as a criticality indicator, more than just a fitting parameter. 

%Keywords: tetrapod nanocrystals; electronic properties; nonlinear optical response; pressure and temperature effects; donor impurity
\end{abstract}

\maketitle

\section{Introduction}\label{intro}

Earthquakes result from the sudden release of accumulated strain along tectonic faults, a process governed by stick-slip dynamics and heterogeneous stress distributions. The inherent complexity of fault systems—characterized by fractal geometries, scale-invariant stress correlations, and nonlinear interactions—makes deterministic earthquake prediction an elusive goal. Instead, statistical and thermodynamic approaches have emerged as powerful tools for understanding seismic phenomena.

Among these, Non-Extensive Statistical Mechanics (NESM), introduced by Tsallis (1988) \cite{Tsallis1988,Tsallis2011}, provides a robust framework for describing systems with long-range interactions, multifractality, and strong correlations—features intrinsic to seismogenesis. The fragment-asperity model, which conceptualizes fault zones as a collection of irregular, stress-bearing fragments (asperities) embedded in a viscoelastic matrix, aligns naturally with NESM, offering a way to describe the statistical distribution of earthquake magnitudes and energy release.

The analysis of seismic events using NESM, particularly Tsallis entropy, has provided significant insights into the complex dynamics of earthquakes. Several studies have demonstrated the presence of non-extensivity in earthquake magnitude and inter-event time distributions, supporting the applicability of Tsallis statistics in seismology.

Silva \textit{et al}. \cite{Silva2006} and Vilar \textit{et al}. \cite{Vilar2007} were among the first to apply NESM to seismic catalogs, showing that earthquake magnitude distributions deviate from the traditional Gutenberg-Richter law and are better described by a $q$-exponential distribution. In subsequent years, several authors have reported on the NESM treatment in relation with different tectonic seismic events throughout the world \cite{Darooneh2009,Telesca2010-1,Telesca2010-2,Telesca2010-3,Vallianatos2010,Vallianatos2012,Telesca2011,Papadakis2015,Barahona2024,Sigalotti2023,Sigalotti2024}.

\begin{figurehere}
	\centering	\includegraphics[width=1\columnwidth,angle=0]{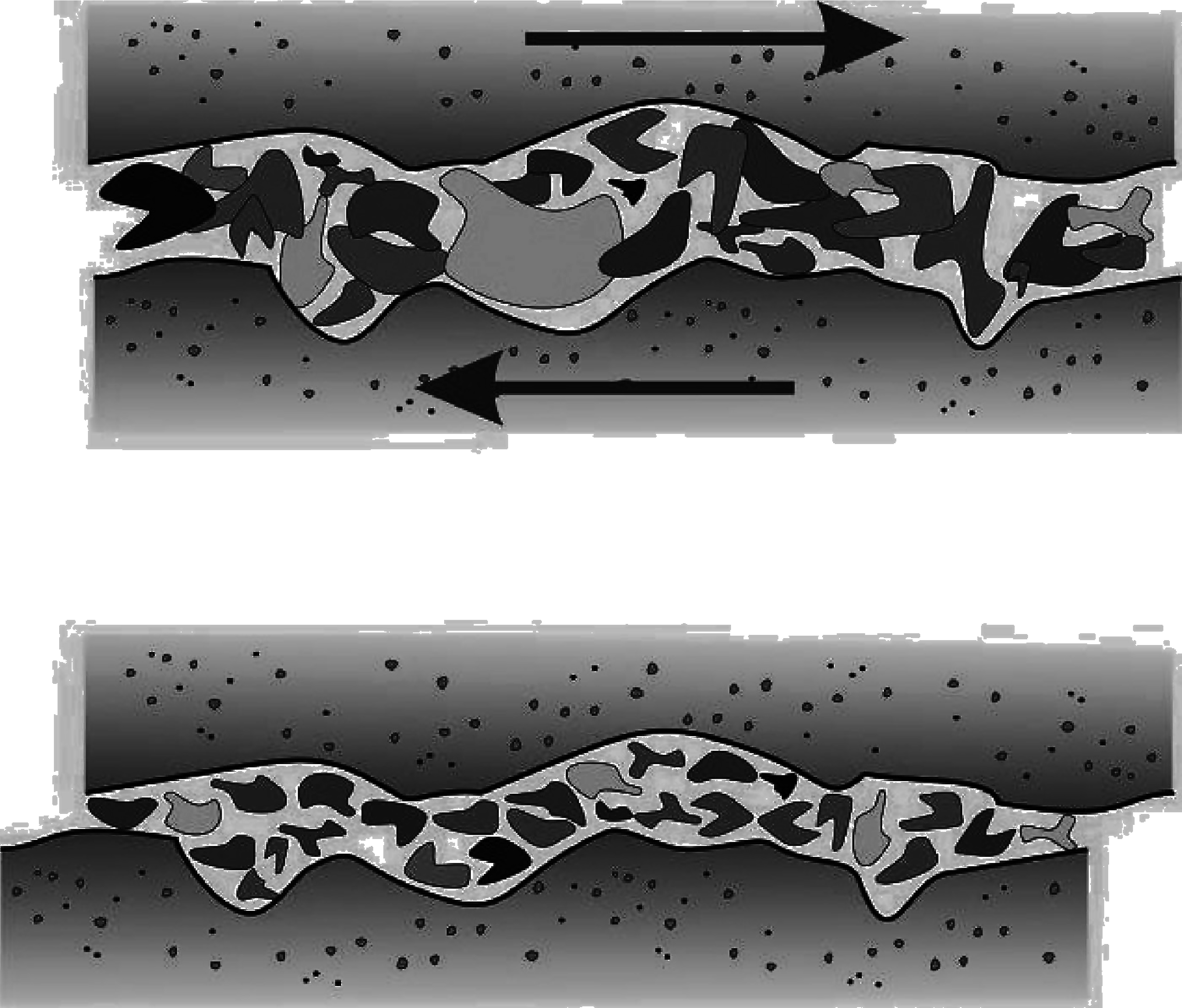}
	\caption{{\small (Top panel - Pre-seismic phase) Before an earthquake occurs, the system exhibits heterogeneous stress distribution and a broad spectrum of fragment sizes. (Bottom panel - Co-seismic phase) The earthquake rupture process leads to fragmentation of asperities and breakdown of barriers, resulting in size homogenization of the fractured materials.}}
	\label{Fig1}
\end{figurehere}

As pointed out by Barahona C\'ardenas and Araujo Soria, the complex system under study also encompasses all faults in the Earth's crust, which themselves exhibit intricate structural properties. By applying nonextensive statistical analysis to subduction zones, we can account for the diverse mechanisms driving seismic activity. Such systems are characterized by long-range spatio-temporal correlations, non-Markovian processes (indicating long-term memory effects), additive and multiplicative noise in mesoscopic Langevin-type equations, weak chaos with near-zero Lyapunov exponents, multifractal geometry, and long-range interactions in many-body systems. Additionally, they often involve nonlinear and/or inhomogeneous mesoscopic Fokker-Planck equations \cite{Barahona2024}. 

Within this context, the fragment-asperity model considers this complexity and describes the tectonic plates having numerous surface asperities with variable size and a whole distribution of fragments accommodating in between. It can be said that all asperities interact to each other via the stress propagation through the fragments. Therefore, this interaction has long range nature. In the relative motion of subduction certain asperities can break, giving origin to a seismic event. Sotolongo and Posadas put forward this idea in 2004 \cite{Sotolongo2004}, and extended it in 2023 \cite{Posadas2023}. Their proposal allowed to modify the Gutenberg-Richter (GR) law into one that includes the $q$-parameter. This makes possible to determine $q$ from the analysis of earthquake catalogs. Examples of the most recent studies along these lines are the references \cite{Barahona2024,Sigalotti2023,Sigalotti2024}. 

Essentially, the fragment-asperity model relies on the following physical picture: Before an earthquake occurs, the system exhibits heterogeneous stress distribution and a broad spectrum of fragment sizes. This configuration allows for numerous possible microscopic configurations ("microstates"), corresponding to a state of relatively high entropy. Then, the earthquake rupture process leads to fragmentation of asperities and breakdown of barriers, resulting in size homogenization of the fractured materials. This reduction in geometric variability decreases the number of allowable microstates, causing a sudden drop in entropy. The entropy reduction occurs rapidly during the dynamic rupture phase, followed by gradual recovery as tectonic stresses begin to rebuild in the post-seismic period \cite{Posadas2023}. A schematic representation of the fragment-asperity situation in a tectonic seismic phenomenon appears in Fig. 1.

In this work, we provide an explanation for the above-mentioned interval of criticality for the $q$-parameter reported in many situations linked to tectonic earthquakes. This is done within the fragment-asperity model by making the novel assumption that the energy is, actually, proportional to the surface of interacting elements (asperities and fragments). Then, the calculation of total entropy, $S$, leads to a functional dependence of this quantity versus $q$ in which it is straightforward to identify the role of this parameter by maximizing $S(q)$. We are also aimed at discussing the possible role of $q$-value as a useful tool in accessing seismic hazard.

\section{Theoretical Framework}\label{theory}

%The physical reasons for the increase and ulterior decrease of entropy in the pre- and post-seismic phases of a tectonic earthquake are those given in the caption of Fig. 1. As it is known, entropy has to do with the multiplicity of "micro" configurations. It tends to be maximal when there are many more ways of stress contact between fragments and plate asperities, which associates with a non-uniform distribution of fragment sizes.

The main assumption within the original non-extensive treatment of the fragment-asperity model \cite{Sotolongo2004} is that the released seismic energy, $\varepsilon$ is directly related to the size of the fragments that occupy the space between fault blocks in tectonic plates: $\varepsilon\sim r$. The revision made by Silva \textit{et al.} put forward a volume-related relationship $\varepsilon\sim r^3$, and a consequent statistical treatment and its relation with G-R description was given  \cite{Silva2006}. The same treatment is adopted by Telesca, who considers the relation $\sigma\sim\varepsilon^{2/3}$ linking the area of fragments and the released energy to derive an expression relating $q$ and the G-R parameter $b$ \cite{Telesca2011,Telesca2012}.

However, in the present study we propose that the seismic energy is, actually, proportional to the contact surface of the fragments: $\varepsilon\sim \sigma$. This assumption is physically grounded in the classical Griffith theory of fracture \cite{Griffith} and the subsequent applications of ideas based on surface energy, brittle rupture and interfacial contacts to the case of fragments that displace with mutual friction \cite{Bowden1950,Bowden1964,Dieterich1979,Ruina1983,Rice1983}. According to that theory, the energy balance governing crack nucleation and propagation is dominated by competition between the elastic energy released in the volume and the surface energy required to create new fracture interfaces. At the breaking threshold, the main energetic cost is associated with the creation and evolution of surfaces, implying that energy dissipation is essentially interfacial rather than volumetric. Concepts along the same lines have been relevant to establish a relation between asperities, surface creation and energy dissipation in fault zones, explaining friction and seismicity as interfacial processes (see a broader discussion in the book by Scholz \cite{Scholz2019}) .

Within the fragment–asperity model framework, the accumulation and release of stress occur through a network of contacts between fragments and fault asperities, such that the relevant microscopic configurations are determined by the total effective contact area. For this reason, the statistical weight of the microstates is naturally expressed as a function of surface rather than volume. By observing the schematic representation in the upper panel Fig. 1, one may realize that strain in the system accumulates not only via the contact sites between the very fragments but also through those between fragments and plate asperities. From a statistical point of view, each different configuration can be considered as a compatible microstates" \cite{Posadas2023}. As it is known, entropy has to do with the multiplicity of these "micro" configurations. It tends to be maximal when there are many more ways of stress contact between fragments and plate asperities, which associates with a non-uniform distribution of fragment sizes. Clearly, such a contact directly relates with the surface of the fragment that undergoes the stress from its surroundings. Therefore, one may construct the probability distribution as a function of the area $\sigma$. 

In a continuous view, the reduced ($k=1$) Tsallis entropy reads as \cite{Tsallis1988,Tsallis2011}

\begin{equation}
	S=\frac{1-\int_{0}^{\infty}p^q(\sigma)d\sigma}{q-1},
\end{equation}

\noindent having the normalization constraints,

\begin{equation}
\int_{0}^{\infty}p(\sigma)d\sigma=1,
\end{equation}

\noindent and $q$-mean of the distribution, 

\begin{equation}
\int_{0}^{\infty}\sigma p^q(\sigma)d\sigma=\langle\langle \sigma\rangle\rangle_q <\infty,
\end{equation}

\noindent is maximized to give the fragment size distribution

\begin{equation}
	p(\sigma)=\frac{(2-q)^{\frac{1}{2-q}}}{\left[1+(q-1)(2-q)^{\frac{q-1}{2-q}}\sigma\right]^{\frac{1}{q-1}}}.
	\label{cuatro}
\end{equation}

Now, we calculate the total entropy by integrating in (1) with respect to the surface $\sigma$, using (4). This gives

\begin{equation}
	S(q)=\frac{1-(2-q)^{\frac{1}{2-q}}}{q-1}.
\end{equation}

It is of particular interest to represent the plots of $S(q)$ and its derivative with respect to $q$. They are shown in Fig. 2.

%Fig. 2
\begin{widetext}
\begin{figurehere}
	\centering
	\includegraphics[width=0.45\columnwidth,angle=0]{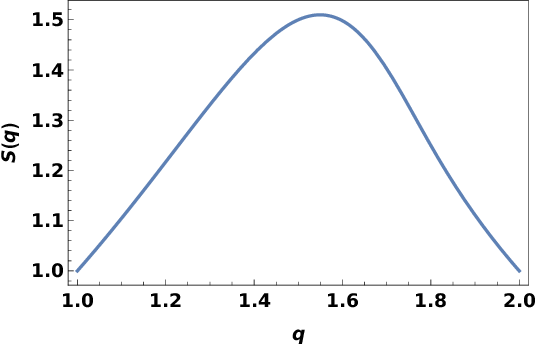}
	\includegraphics[width=0.45\columnwidth,angle=0]{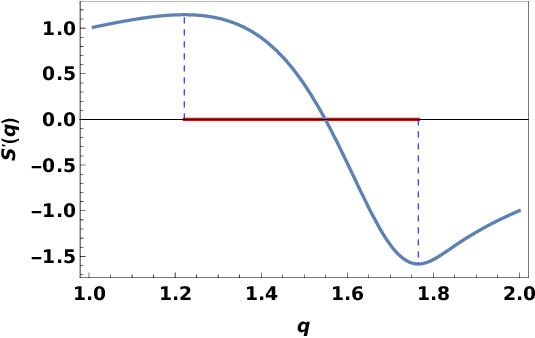}	
	\includegraphics[width=0.45\columnwidth,angle=0]{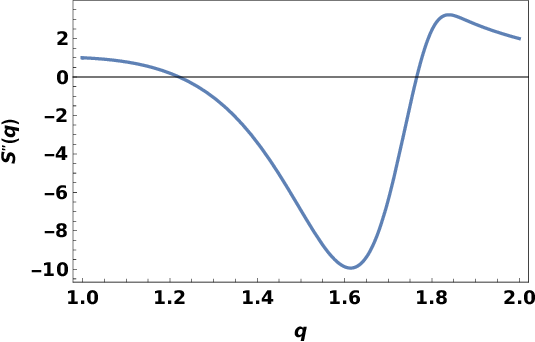}	
	\caption{(Top left) Total non-extensive (reduced) entropy, $S(q)$, of the fragment-asperity model as a function of the $q$-parameter. (Top right) The first derivative of $S(q)$, as a function of the $q$-parameter. (Bottom) The second derivative of $S(q)$, as a function of the $q$-parameter. The marked horizontal interval extends between the first ($q=1.22$) and second ($q=1.76$) inflection points of the entropy. It is readily apparent that the rate of change of $S(q)$ is more pronounced in the range of $1.4\leq q\leq 1.75$. The maximum of $S(q)$ lies at $q=1.549$. The minimum of $S^{\prime\prime}(q)$ lies at $q=1.614$. }
    \label{Fig2}
\end{figurehere}
\end{widetext}

Within the fragment-asperity framework, the energetic scaling is fixed by the Griffith fracture relation $\varepsilon\propto\sigma$. However, as put forward by Lyra and Tsallis \cite{Lyra1998} the non-extensive parameter $q$ does not encode energetic effects but rather the geometrical correlations associated with the fractal distribution -in this case- of asperities and fragments along the plate contact surface.

From the mathematical point of view, the linear relation between dissipated energy and surface area reveals as the only one that leads to a closed elementary form for a well-defined total entropy as a function of $q$. In the Appendix A we present the proof for this assertion.

Looking to clarify the interplay between energetic and geometric aspects in the model, we aim at discussing the implication of our the above mentioned relation by deriving a quantitative expression that allows to determine the value of non-extensivity parameter from actual seismicity reports. This is the one between $q$ and the $b$ parameter of G-R law for the logarithm of cumulative earthquake number versus $M$, within the intermediate magnitude regime: $Log_{10}N_c(N>M)=a-bM$. 

First, it is worth pointing out that, with this assumption, eqn. (4) leads to a probability distribution, $p(\varepsilon)$, that depends linearly on the energy variable in its argument. This contrasts with the dependencies on $\varepsilon^2$ and $\varepsilon^{2/3}$ found in the works of Refs. \cite{Sotolongo2004} and \cite{Silva2006,Telesca2012}, respectively, with their corresponding proposals for $\varepsilon(\sigma)$. This fact is relevant because it consequently determines the shape of $b(q)$ and, ultimately, the parameter that governs the non extensive thermodynamic description of a given seismic region. 

We have followed the same path of reasoning described in Refs. \cite{Silva2006,Telesca2012} to obtain the corresponding formula $b(q)$. Our result is: 

\begin{equation}
	b(q)=\frac{3}{2}\left ( \frac{2-q}{1-q}\right ). 
\end{equation}

However, a simpler and more intuitive procedure would lead to the same result. We refer the reader to Appendix B below for details.  

Within this scheme, the presence of the $3/2$ factor in the relationship between G-R $b$-value and the non-extensivity parameter $q$ differs from the factors $2$ and $1$ obtained with the mentioned dependencies of $p(\varepsilon)$ on $\varepsilon^2$ and $\varepsilon^{2/3}$, respectively. In our opinion, its appearance can be interpreted as a direct consequence of the geometric scaling between surface and volume in three-dimensional fragmentation processes. While the energy associated with each fragment scales with its surface area, the cumulative number of events is controlled by volumetric statistics. The ratio between these two scaling laws naturally leads to the three-halves factor, thereby providing a geometric and physical interpretation linking nonextensive statistics, fracture mechanics, and the release of seismic energy.

\section{Results and discussion}\label{results}

%\textcolor{blue}{Los puntos de inflexión se encuentran en $q=1.22128$ y $q=1.7625$. El máximo de la entropía se ubica en $q=1.54921$.}

We aim to analyze the outcome of this approach considering the different reported values of the non-extensivity parameter in studies applying NESM to the seismic catalogs referenced in the introduction. For instance, in the early works of Refs. \cite{Silva2006,Vilar2007}, $q$-parameter values lie typically in the range $1.6 \leq q \leq 1.7$, indicating strong non-extensive behavior.

Further studies by Darooneh and Mehri \cite{Darooneh2009} and Telesca \cite{Telesca2010-1,Telesca2010-2,Telesca2010-3} reinforced these findings. Telesca \cite{Telesca2010-1} examined Italian seismicity, finding $q \approx 1.6–1.8$, suggesting long-range interactions and memory effects in earthquake dynamics. In another study \cite{Telesca2010-2}, he analyzed seismic sequences in different tectonic settings, obtaining $q \approx 1.5–1.9$, with higher $q$-values correlating with more complex fault systems.

Vallianatos and Sammonds \cite{Vallianatos2010} and Vallianatos \textit{et al}. \cite{Vallianatos2012} expanded this framework to different regions, including the San Andreas Fault, reporting $q \approx 1.6–1.8$, consistent with criticality in earthquake preparation processes. Telesca further confirmed these results, showing that the $q$-parameter varies with tectonic activity, with higher values ($q > 1.7$) in regions of intense deformation \cite{Telesca2011}.

These studies collectively demonstrate that seismic phenomena exhibit non-extensive behavior, characterized by $q$-values consistently greater than 1 (the Boltzmann-Gibbs limit), reflecting the presence of long-range correlations, multifractality, and scale-invariant properties in earthquake dynamics. The $q$-parameter thus serves as a key indicator of the degree of non extensivity in seismic systems, providing deeper understanding of their underlying physics. Reports about the increase of $q$ previous the occurrence of earthquakes of high magnitude have appeared in recent times \cite{Papadakis2015,Barahona2024}.

Clearly, the range of values of $q$-parameter identified in the vast majority of cases coincides with the interval of most pronounced variation in Tsallis entropy within the fragment-asperity model, as can be seen in the right panel of Fig. 2. This interval includes the maximum value of this quantity as a function of q, which is $q_m = 1.55$.

If we consider, for example, the most recent reports (2024), we find that in the analysis of seismic activity along Mexico's western coastline [recorded in the 2000-2023 catalog of Mexican National Seismological Service (MNSS)], the general result for the non-extensivity parameter is $1.52 \lesssim q \lesssim 1.61$. This activity is primarily associated with tectonic plate interactions along the southern Pacific coast as well as active structures in the Gulf of California, and the reported values of $q$ arise from fitting of the non-extensive normalized cumulative distribution reported in \cite{Silva2006}, using quite sophisticated optimization algorithms \cite{Sigalotti2024}. Meanwhile, the study of data from Ecuador's subduction zone (1901-2021) shows that the occurrence of $M > 5$ earthquakes is associated with values of $1.68 \leq q\leq 1.75$, with noticeable decreases in this parameter following such events \cite{Barahona2024}. One interesting fact revealed in this latter study is the relation of $q$ versus time around the date of the $M=7.8$ Manab\'i earthquake of April 2016: A sudden increase just before the event, followed by a pronounced fall after it. This fact coincides with the observations of Papadakis \textit{et al}. for the Kobe earthquake of 1995 \cite{Papadakis2015} and Telesca for the L'Aquila one in 2009 \cite{Telesca2010-3}. 

To continue with the verification process, we have applied expression (6) to evaluate $q$ using the values of $b$ resulting from the fitting of the cumulative distributions associated to earthquakes contained in the above mentioned MNSS catalog, including information on more that 190 thousand seismic events registered for six regions along the west coastline of Mexico \cite{Sigalotti2024}. These $b$ values and the resulting -appropriately rounded- $q$ from our formula appear in Table I.

In addition, we performed a second analogous comparison in order to produce another comparative analysis. In that case, the cumulative number, $N_c$, of earthquakes with magnitude above or equal to a given $M_{min}$ has been calculated for the set of earthquake reports belonging to the following seismic regions: (a) The Southwestern region of South America, with $M_{min}= 3$; (b) the region extending from Southern California to Vancouver island along the Pacific coast, including the so-called Cascadia subduction zone; (c) a revisiting of the Southwestern region of Mexico, limited from south of L\'azaro C\'ardenas (state of Michoac\'an) to southwest of Tapachula (state of Chiapas); (d) the Kamchatka peninsula. Data were taken from the catalog posted on the website of U. S. Geological Survey \cite{ChileanQuakes}. In all cases, the events expand within a time period that encompasses from January 2020 to mid July 2025, with the exception of Mexican data that extend to the first half of January 2026 in order to include the very recent earthquake with $M=6.5$ in San Marcos  (state of Guerrero). All data are provided in the supplementary material. Figure \ref{Fig4} contains the resulting plot of $Log_{10}N_c$ versus the magnitude for each case, including the linear fit of G-R type that expands through the intermediate range of magnitudes. The use of expression (6) produces the following values of the non-extensivity parameter: (a) $q=1.588$, (c), and (d) $q=1.590$; (b) $q=1.649$.

Three of the calculated values of $q$ are within a close range around $q_m$. This points at a relation of tectonic earthquake activity and the non-extensive configuration that corresponds to maximal total Tsallis entropy of the fragment-surface complex system. Notably, they all have in common the occurrence of giant tectonic-type earthquakes with $M\geq 8$ within the period, together with tens of events with $M\geq 6$. 

In contrast, although its value of $q$ lies inside the interval of strong variation of $S(q)$ highlighted in Fig. 2, the result for case (b) departs toward the region where $S^{\prime\prime}(q)$ has its fastest growing, above $q=1.614$ whilst the other three are in the interval of fastest descend of the total entropy, at the left of that value. This seismic zone presents a distinct behavior within the fragment–asperity formalism. It is readily apparent that the logarithmic cumulative distribution for this region does not display the same shape as in the other three cases. Notably, it does not show giant seismic events during the time interval analyzed, with the strongest peaking at $M=7.2$ in June 2005, and having only 16 of them with $M\geq 6$.  Its cumulative distribution $Log_{10}N_c$ also departs from the quasi-linear G–R behavior observed in the other cases, and the Gutenberg–Richter fit for this region yields $b = –0.81$, which, through the relation (6) corresponds to the above mentioned value $q = 1.649$. This value clearly differs from those obtained for the other regions, which cluster around $q \approx 1.59$, close to the maximum of the total entropy $S(q)$. In contrast, the California–Vancouver result lies on the right-hand side of the interval between inflection points, in the sector where the second derivative $S^{\prime\prime}(q)$ increases more rapidly.

Within the thermodynamic interpretation adopted here, this configuration that shifts the q-value toward the region of growing curvature of $S(q)$ may indicate that variations in seismic activity produce relatively larger changes in the complexity of the fragment–surface ensemble. Here, a possible explanation could be the actual presence of a different underlying physics (transform faults?) that modifies the multifractal geometry of the fractured seismic region beyond the one related with plate subduction. Thus, the elevated $q$ could be signaling a different non-extensive regime, consistent with the absence of large-magnitude events and with the structural properties of the regional fault system. Considering this hypothesis, a detailed analysis of this particular region could be performed elsewhere.

In spite of the particularity of region (b), in the analyzed cases there are many elements of agreement that significantly correlate with what we see in Fig. 2 for $S(q)$ and $S^{\prime}(q)$. As a function of the complexity, represented by the $q$-parameter, the interval of highest entropy values and of its steepest variation, coincides with the range arising from the analysis of plate subduction seismicity around the world -largely in relation with mainshocks ($M>5$). This mathematical feature speaks about the criticality related with tectonic activity, described using NESM within the fragment-asperity model. Also, it points at confirming the long-range nature of the involved interactions as well as the need for a non-extensive description of these phenomena.

\begin{widetext}
\begin{center}
\begin{table}
\caption{Gutenberg-Richter $b$-parameter for earthquake events from six seismic regions along the west coastline of Mexico (data taken from Ref. \cite{Sigalotti2024}, see also their Figure 3) and the non-extensive $q$, calculated from equation (6). }
\begin{tabular}{c|c|c|c}
	\noalign{\hrule height 1.5pt}
	%\hline
	Region & Number of Earthquakes & $\;\;\;\;\;\;$b$\;\;\;\;\;\;$ & $\;\;\;\;\;\;$q$\;\;\;\;\;\;$ \\
	\hline
	Baja California & 9579 & -0.991 &  1.60\\
	%\hline
	Nayarit-Jalisco & 3774 & -1.135 &  1.57\\
	%\hline
	Colima-Michoac\'an & 20229 & -1.148  &  1.57\\
	%\hline
	Guerrero & 33080 & -1.243 &  1.55\\
	%\hline
	Oaxaca & 59803 & -1.327 &  1.53 \\
	%\hline
	Chiapas & 67008 & -1.359 &  1.52 \\
	%\hline
	\noalign{\hrule height 1.5pt}
\end{tabular}
\end{table}
\end{center}
\end{widetext}

\begin{widetext}
\begin{figurehere}
	\centering	
	\includegraphics[width=0.45\columnwidth,angle=0]{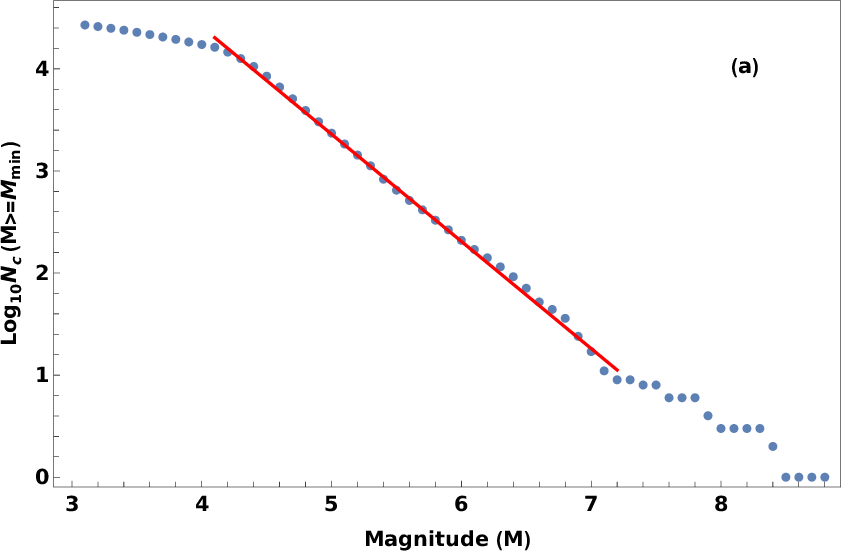}
	\includegraphics[width=0.45\columnwidth,angle=0]{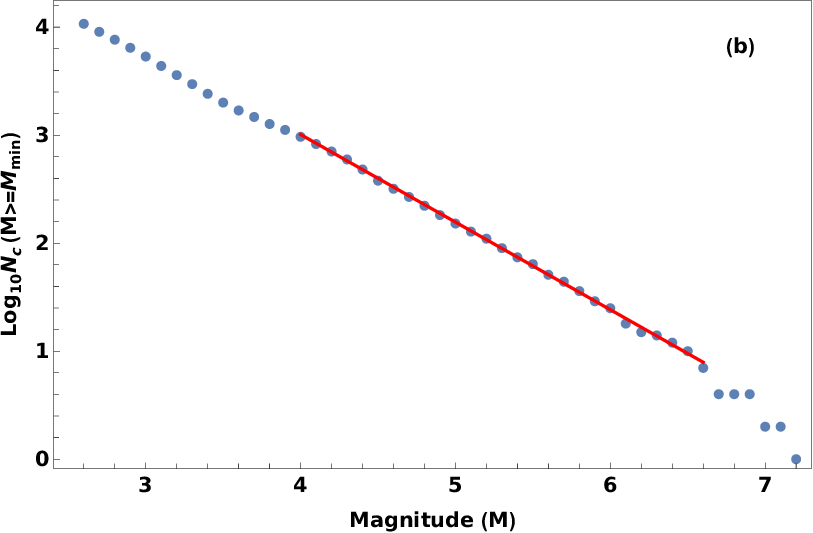}
	\includegraphics[width=0.45\columnwidth,angle=0]{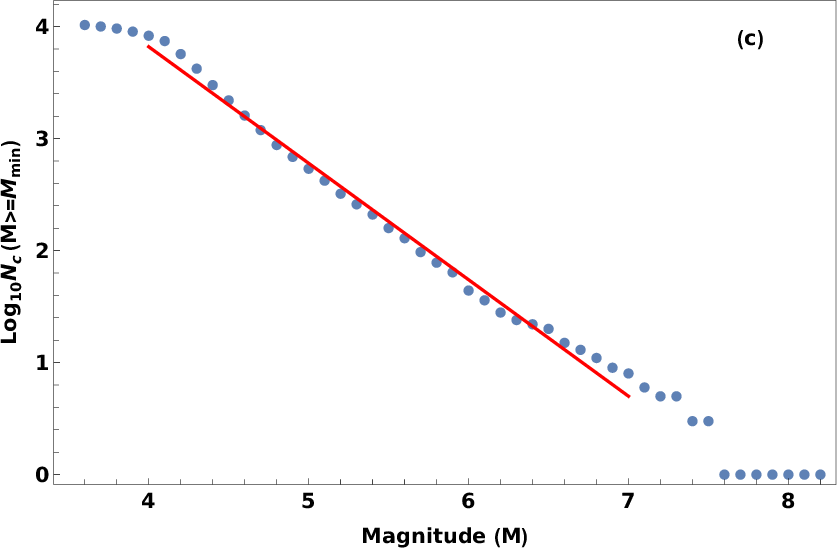}
	\includegraphics[width=0.45\columnwidth,angle=0]{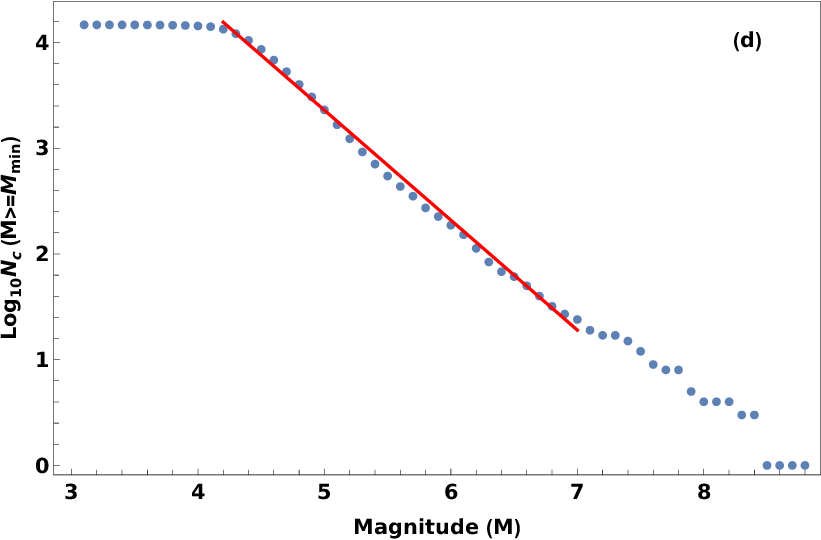}
	\caption{{\small The plot of the logarithmic cumulative number of seismic events versus earthquake magnitude corresponding to (a) Western South America region; (b) The Western region of North America from Southern California to Vancouver Island; (c) The Southwest of Mexico, and (d) Kamchatka region. Data encompasses events from January 2020 to July 2025, except case (c) for which the period extends to January 2026, to include the more recent earthquake registered in the country with $M=6.5$. (https://earthquake.usgs.gov/earthquakes. Details of every seismic event considered for each region are provided in the CSV file attached as supplementary material). The straight line corresponds to the linear Gutenberg-Richter fit along the intermediate $M$ values, with slope: (a) $b=-1.05$; (b) $b=-0.81$; (c) and (d) $b=-1.04$.}}
	\label{Fig4}
\end{figurehere}
\end{widetext}

In our opinion, to identify the role of $q$ in the properties of total Tsallis entropy as a measure of the microstates distribution in the statistical approach to tectonic seismicity becomes a conceptual breakthrough that strengthens the fragment-asperity model and definitively removes this number from the category of fitting parameter, as it has been considered in previous works. At the same time, our analysis theoretically supports considering $q$ as an indicator of criticality.

First and foremost, by assuming the direct dependence of seismic energy on the contact surface between fragments and asperities we have obtained a suitable tool to identify a range of critical values when dealing with the seismic dynamics: the total Tsallis entropy $S(q)$ [eqn (5)]. Moreover, the empirical interval discussed coincides with the region of maximal sensitivity for the entropy of the non-extensive ensemble. 

The $q$-parameter acquires a more concrete physical role: rather than being solely a measure of departure from extensivity, it can also provide complementary information about the evolving seismic state of a region. Current methodologies allow the characterization of seismic regions with major events by comparing their $q$-values with preceding $q$-sequences, thereby offering a potential tool for hazard assessment. Importantly, neither the value of $q$ nor its temporal evolution is sufficient on its own to forecast the exact timing or location of future seismic events. What $q$ offers is a statistical descriptor whose critical ranges or anomalous fluctuations deserve careful investigation. In this regard, the systematic analysis of $q$-sequences derived from seismic catalogs may help enrich the broader framework of seismic monitoring.

\section{Conclusions}\label{conclusions}

The present work provides a thermodynamic interpretation of seismic non-extensivity. The possibility of evaluating the total Tsallis entropy as a function of the non-extensivity parameter for the fragment-asperity model allows to identify the interval of $q$-values where this quantity exhibits its more pronounced variation, including its maximum value. Such interval includes the ranges determined for $q$ in all investigated NESM approaches of tectonic seismic activity around the world. This fact directly speaks about the complexity of this natural phenomenon, highlights the particular role of $q$ beyond its use as a fitting parameter. 

We have shown that the linearity $\varepsilon\propto\sigma$ is not a a convenient or empirical choice: it is the only functional relation that produces a total entropy, $S(q)$, with a closed, elementary, form, without special functions or artificial cutoffs. This structural mathematical uniqueness, combined with its physical grounds from Griffith-Barenblatt theory on surface energy in fracture processes, likely makes this relation the only suitable scaling  consistent with a well-defined statistical mechanics of the fragment-asperity model. In this formulation, the parameter $q$ reflects the geometrical correlations of the system, while the energetic contribution is entirely determined by the Griffith surface-energy relation.

By considering the direct proportionality between the surface area and the dissipated energy as a key assumption, our formulation directly leads to a relation between G-R, $b$, and the non-extensivity, $q$, parameters which allows for the evaluation of the latter from the analysis of the cumulative distribution of events in a given seismic region. Although that is something already made in previous reports, the particularity in this case is that eqn. (6) arises in a natural way when the derived Tsallis probability and empirical G-R relations are properly related. Moreover. the factor involved in that expression differs from the other reports in such a way that points at an intrinsic geometric relation between the scales of the counted magnitudes: energy and cumulative number of events.

As a follow up, the controversial case of study about fracking process could be investigated through $q$-value dynamics in active extraction zones. Such analysis could disentangle the relative contributions of different mechanisms (e.g. plate fracturing, fluid injection), quantify their environmental impacts, particularly regarding induced seismicity and establish causal hierarchies among triggering factors. A manuscript addressing this specific application is currently in preparation.

\appendix

\section{Uniqueness of the relation $\varepsilon \propto \sigma$ for a well-defined total entropy}

\textit{Structural Uniqueness Theorem}: Let $f:\mathbb{R}^+\rightarrow \mathbb{R}^+$ be a monotonically increasing, differentiable function with $f(0)=0$ that relates the energy $\varepsilon$ to the contact area $\sigma$ through $\varepsilon = f(\sigma)$. Then, the total Tsallis entropy

\begin{equation}
	S(q)=\frac{1-\int_0^{\infty}p^q(\sigma)d\sigma}{q-1}
\end{equation}

\noindent is a well-defined elementary function (free of \textit{ad hoc} cutoff parameters and special functions) if and only if $f(\sigma)=k\sigma$, with $k>0$ constant.

\textit{Proof}: The maximization of the entropy under the conditions given in (2) and (3) yields a $q$-exponential distribution

\begin{equation}
	p(\sigma)=A\left[1+Bf(\sigma) \right]^{-\frac{1}{1-q}},
\end{equation}

\noindent where $A$ and $B$ are normalization constants depending on $q$, and where we have explicitly assumed that the distribution depends on the energy through $\varepsilon=f(\sigma)$.

The total entropy requires evaluating the integral

\begin{equation}
	I(q)=\int_0^{\infty}p^q(\sigma)d\sigma
	= A^q\int_0^{\infty}\left[1+Bf(\sigma) \right]^{-\frac{q}{1-q}}d\sigma.
	\label{AB}
\end{equation}

\noindent For $S(q)$ to be an elementary function of $q$, this integral must be analytically solvable without leaving unevaluated integrals or introducing artificial upper limits of integration, $\sigma_{max}$. Let us therefore perform the change of variables $u=Bf(\sigma)$, yielding:

\begin{equation}
	\sigma=f^{-1}(u/B);\;\;\; d\sigma=\frac{1}{B}\frac{1}{f^{\prime}\left[f^{-1}(u/B)\right]}du,
\end{equation}

\noindent and thus the integral becomes:

\begin{equation}
	I(q)=\frac{A^q}{B}\int_0^{\infty}(1+u)^{-\frac{q}{1-q}}
	\frac{1}{f^{\prime}\left[f^{-1}(u/B)\right]}du.
\end{equation}

We now seek the condition for a closed elementary form. This is achieved if the factor

\begin{equation}
	g(u)=\frac{1}{f^{\prime}\left[f^{-1}(u/B)\right]}
\end{equation}

\noindent is a constant, independent of $u$. Indeed, if we require

\begin{equation}
	\frac{1}{f^{\prime}\left[f^{-1}(u/B)\right]}=\frac{1}{C}= \text{constant},
	\label{BB}
\end{equation}
and denote $x=f^{-1}(u/B)$, then $u=Bf(x)$ and condition \ref{BB} becomes $f^{\prime}(x)=\text{constant}$ for all $x>0$. The only function whose derivative is constant is the linear function: $f(x)=kx+d$, and with the condition $f(0)=0$ (zero energy when the area is zero), we obtain $d \equiv 0$. Consequently,

\begin{equation}
	f(\sigma)=k\sigma,
\end{equation}
with which $g(u)=1/k$, and the integral \ref{AB} reduces to

\begin{equation}
	I(q)=\frac{A^q}{Bk}\int_0^{\infty}(1+u)^{-\frac{q}{1-q}}du,
\end{equation}
which is a standard Beta integral with closed-form solution

\begin{equation}
	I(q)=\frac{A^q}{Bk}(q-1).
\end{equation}

Substituting $A$ and $B$ as given in \ref{cuatro} finally leads to

\begin{equation}
	S(q)=\frac{1-(2-q)^{\frac{1}{2-q}}}{q-1}
\end{equation}
as discussed in the main text.

If $g(u)$ is a non-constant function of $u$, the integral \ref{AB} generally lacks a closed elementary form. For example, in the simplest case $g(u)=u^a$ with $a$ different from zero—but $-1<a<1/(q-1)$ to ensure convergence—the result of the integral is the Beta function $B[1+a,-a+1/(q-1)]$. Beyond this example, one encounters other special functions such as the incomplete Gamma function or hypergeometric functions; or situations in which it is necessary to introduce an integration cutoff at $\sigma_{max}$ or to carry out numerical computations for each value of $q$. In all such cases, the total entropy $S(q)$ loses its character as a well-defined elementary function, and the thermodynamic analysis of the system becomes problematic.

\section{Alternative derivation of the $b(q)$ relation}

Taking into account the direct proportionality between dissipated energy and surface area, we can make use of expression (4) to evaluate the cumulative distribution

\begin{equation}
	N_c(>\varepsilon)=\int_\varepsilon^{\infty} p(\varepsilon^{\prime})d\varepsilon^{\prime}.
\end{equation}

For $q>1$, the asymptotic behavior of this quantity is a well-known result for type $q$-exponential distributions;

\begin{equation}
	N_c(>\varepsilon)\propto \varepsilon^{-\frac{2-q}{q-1}}.
\end{equation}
	
Now, the G-R law gives:

\begin{equation}
	N_c(>M)\propto 10^{-bM},
\end{equation}

\noindent which, together with the generally accepted (although not free of dispersion) empirical relation \cite{Kanamori1978}

\begin{equation}
	M\sim \frac{2}{3}Log_{10}\varepsilon\;\implies \varepsilon\sim 10^{\frac{3}{2}M},
\end{equation}

\noindent allows to write the G-R law in terms of energy:

\begin{equation}
	N_c(>\varepsilon)\propto \varepsilon^{-\frac{2}{3}b}
\end{equation}

\noindent and a comparison between the two above expressions for $N_c(>\varepsilon)$ yields the result of eqn. (6).

\bibliography{bibliography-Sotolongo}
\end{document}